\documentclass[onecollarge]{svjour2}
\usepackage{graphicx,amsmath,amsbsy,amssymb,float}

\def\la{\langle}
\def\ra{\rangle}
\def\ve{\vert}
\def \d{\mathrm{d}}

\journalname{J Stat Phys}

\begin{document}

\title{Statistical Mechanics of Nucleosomes Constrained by Higher-Order Chromatin Structure
}

\titlerunning{Statistical Mechanics of Nucleosomes}        


\author{R\u{a}zvan~V.~Chereji \and Alexandre~V.~Morozov}

\authorrunning{Chereji, Morozov} 

\institute{R.V. Chereji \at
								Department of Physics and Astronomy, Rutgers University, Piscataway, NJ 08854-8019, USA
								\email{rchereji@physics.rutgers.edu}
              \and
           A.V. Morozov \at   
              	Department of Physics and Astronomy, Rutgers University, Piscataway, NJ 08854-8019, USA\\             
              	BioMaPS Institute for Quantitative Biology, Rutgers University, Piscataway, NJ 08854-8019, USA\\
								\email{morozov@physics.rutgers.edu}
}
 
\date{May 24, 2011}

\maketitle

\begin{abstract}
Eukaryotic DNA is packaged into chromatin: one-dimensional arrays of nucleosomes separated by stretches of linker DNA
are folded into 30-nm chromatin fibers which in turn form higher-order structures \cite{Felsenfeld:2003}.
Each nucleosome, the fundamental unit of chromatin, has 147 base pairs (bp) of DNA wrapped around a histone octamer \cite{Richmond:2003}.
In order to describe how chromatin fiber formation affects nucleosome positioning and energetics, we have developed a thermodynamic model of finite-size particles
with effective nearest-neighbor interactions and arbitrary DNA-binding energies.
We show that both one- and two-body interactions can be accurately extracted from one-particle density profiles based on
high-throughput maps of \textit{in vitro} or \textit{in vivo} nucleosome positions. Although a simpler approach that
neglects two-body interactions (even if they are in fact present in the system) can be used to predict sequence determinants of nucleosome positions,
the full theory is required to disentangle one- and two-body effects.
Finally, we construct a minimal model in which nucleosomes are positioned by steric exclusion and
two-body interactions rather than intrinsic histone-DNA sequence preferences. The model accurately reproduces nucleosome occupancy
patterns observed over transcribed regions in living cells.
\end{abstract}

\section{Introduction} \label{section:Introduction}

Eukaryotic DNA in packaged into nucleosomes. Each nucleosome consists of 
a 147 bp-long DNA segment wrapped around a histone octamer in $\sim 1.7$ turns of a left-handed superhelix \cite{Richmond:2003}.
In addition to its primary function of DNA compaction, chromatin modulates DNA accessibility
to transcription factors and other molecular machines,
affecting gene transcription as well as DNA repair, maintenance, and replication.
In \textit{S. cerevisiae}, arrays of nucleosomes cover $\sim 80 \%$ of genomic DNA, exerting profound influence on gene regulatory
programs \cite{Struhl:1999,Mellor:2006,Li:2007,Bai:2010}.

Because the free energy of bending a DNA segment into a superhelix will vary depending on its nucleotide
sequence and composition \cite{Olson:1998,Morozov:2009}, nucleosomes exhibit a range of \textit{in vitro} formation energies \cite{Lowary:1998,Thastrom:1999}
(although almost any DNA sequence can be packaged into a nucleosome).
Recent work has clarified the role of sequence rules that influence nucleosome positioning: genome-wide \textit{in vitro} reconstitution experiments
have confirmed that nucleosome architecture over promoters and genes is partially established by DNA sequence, mostly as a result of nucleosome depletion
from A/T-rich, nucleosome-disfavoring sequences on both ends of the transcript \cite{Sekinger:2005,Kaplan:2009,Zhang:2009}.
However, even if genomic DNA from \textit{S. cerevisiae} is mixed with histones in a 1:1 mass ratio
(leading to the maximum nucleosome occupancy of $0.82$ which is close to the \textit{in vivo} value \cite{Zhang:2009}),
nucleosomes are not strongly localized and, on average, nucleosome occupancy is just $\sim 20-30 \%$ lower over
nucleosome-depleted regions (NDRs) compared to the mean occupancy in a window which includes both the coding region and adjacent
sequence. The absence of nucleosome localization \textit{in vitro} and shallow NDRs
indicate that the absolute magnitude of intrinsic histone-DNA interactions is less than $1~k_B T$.

\begin{figure}[t]
  \begin{center}
  \includegraphics[width=3.2in]{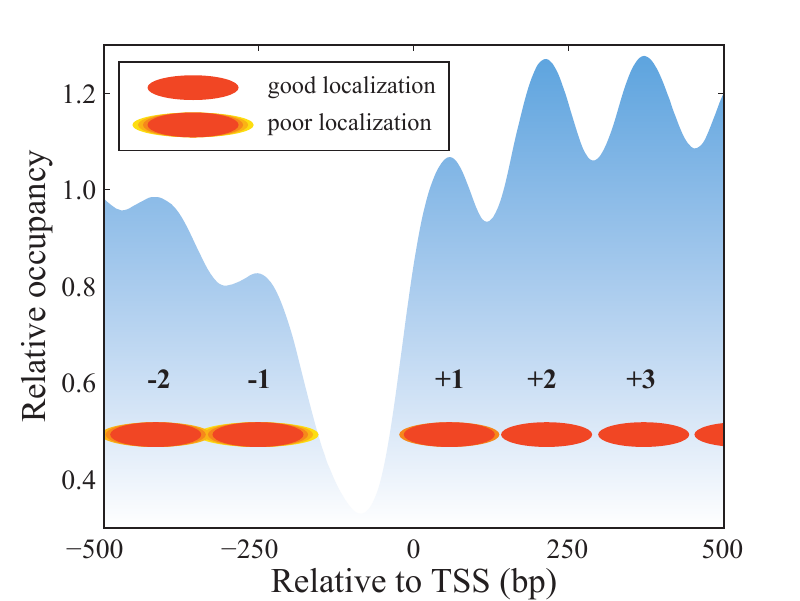}		
  \end{center}
  \caption{(Color online) A typical nucleosome configuration near the transcription start site (TSS).
The blue background represents \textit{in vivo} nucleosome occupancy from Zawadzki et al. \cite{Zawadzki:2009},
averaged over all genes. The occupancy profile is normalized by its average in the $[-500,500]$ bp window.
The intensity of the blue color is proportional to the degree of nucleosome localization.
}
\label{Fig:HardRodsConfig}
\end{figure}

\textit{In vivo}, 5' and 3' NDRs flanking the transcript
are much more pronounced ($\sim 60-70 \%$ occupancy depletion on average with respect to the mean \cite{Kaplan:2009,Yuan:2005,Mavrich:2008a,Mavrich:2008b}),
establishing a striking pattern of nucleosome localization over genic regions simply
due to steric exclusion which causes nucleosomes to ``phase off'' potential barriers \cite{Kornberg:1988} (Fig.~\ref{Fig:HardRodsConfig}).
Although the exact nature of these \textit{in vivo} barriers is unknown and may vary between cell types and environmental conditions, 
they are likely established through a combined action of RNA polymerase, ATP-dependent chromatin remodeling enzymes and DNA-binding
proteins \cite{Workman:2006,Barrera:2006,Schnitzler:2008}.

Nucleosome positions and formation energies can be predicted using a thermodynamic model which
takes intrinsic histone-DNA sequence preferences and one-dimensional steric exclusion into account \cite{Locke:2010}.
In this approach, sequence determinants of nucleosome energetics are inferred directly from experimentally available
nucleosome occupancy profiles. The profiles are obtained by isolating and sequencing mononucleosomal DNA on a large scale, followed
by mapping nucleosomal sequence reads to the reference genome \cite{Tolkunov:2010}.
However, structural regularity of the chromatin fiber imposes additional constraints on nucleosome positions \cite{Ulanovsky:1986,Widom:1992}:
linkers between neighboring nucleosomes become preferentially discretized with the $10-11$ bp periodicity of DNA helical twist \cite{Wang:2008}.
The discretization is required to avoid steric clashes caused by the nucleosome rotating with respect to the linker DNA axis as the linker
increases in length \cite{Ulanovsky:1986},
and more generally to maintain a regular pattern of protein-protein and protein-DNA contacts in the chromatin fiber \cite{Widom:1992}.
Indeed, adding a short DNA segment to the linker will rotate the nucleosome with respect to the rest of the fiber,
causing disruption of its periodic structure. The disruption is minimized if the length of the extra segment is a multiple of $10-11$ bp,
which brings the nucleosome into an equivalent rotational position.

\begin{figure}[t]
	\begin{center}
	\includegraphics[width=\textwidth]{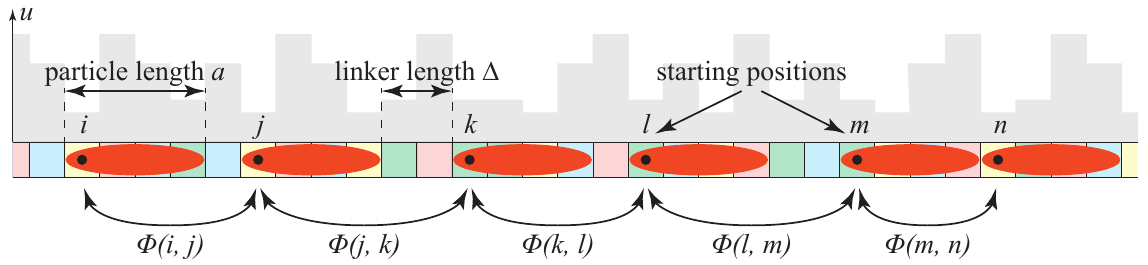}
	\end{center} 
	\caption{(Color online) A typical configuration of $6$ nucleosomes. Each nucleosome covers DNA sequence (represented by colored boxes)
which gives the one-body energy of the nucleosome, $u$. The one-body energy is represented by gray bars. For simplicity, in this toy model,
the one-body energy is entirely determined by the base pair located at the starting position of the nucleosome.
In more realistic scenarios one-body energy is a function of the entire sequence occupied by the nucleosome.
The two-body interaction $\Phi(i,j)$ acts only between neighboring nucleosomes, with two indices $i$ and $j$ representing
their starting positions.}
\label{Fig:NucleosomesSketch}
\end{figure}

We have recently developed a rigorous approach in which linker length discretization is described by
nearest-neighbor two-body interactions in a 
system of non-overlapping finite-size particles \cite{Chereji:2010} (Fig.~\ref{Fig:NucleosomesSketch}). We have shown that 
it is possible to infer one-body energies given by intrinsic
histone-DNA interactions simultaneously with two-body energies caused by chromatin fiber formation. 
The two-body potential can be deduced even in the presence of one-body energies
related to rotational positioning of the nucleosome \cite{Mavrich:2008b,Tolkunov:2010,Zhang:2009},
which have the same $10-11$ bp helical twist periodicity.
We have predicted the two-body interaction from high-throughput maps of nucleosome positions
on the \textit{S. cerevisiae} genome, and demonstrated its essential role in
forming nucleosome occupancy patterns over genic regions.

Here we present a detailed account of our theoretical framework.
We also show that sequence determinants of nucleosome positioning can be obtained with
an interaction-free model \cite{Locke:2010}, even if in reality the two-body potential and histone-DNA interactions
have comparable magnitudes. To this end, we develop a minimally constrained sequence-specific 
model of nucleosome energetics in which the same energies are assigned to mono- and dinucleotides regardless of their exact
position within the 147 bp nucleosomal site \cite{Locke:2010}.
However, only by taking into account the two-body interaction can we make a clear distinction between the two 
types of energies which dictate nucleosome positioning. Finally, we build a minimal model in which \textit{in vivo} nucleosomes 
are positioned solely by potential barriers located at each end of the transcript.
Without invoking explicit sequence specificity, the model successfully reproduces nucleosome occupancy patterns observed \textit{in vivo} in \textit{S. cerevisiae}.
In contrast, sequence-dependent models can only capture liquid-like, delocalized behavior observed with \textit{in vitro}
nucleosomes \cite{Kaplan:2009,Zhang:2009}. By combining the minimal model with sequence-specific nucleosome energies,
we estimate that intrinsic histone-DNA interactions contribute $< 30 \%$ to the height of the \textit{in vivo} potential barriers.

\section{Theory} \label{section:Theory}

\subsection{Energetics of one-dimensional hard rods with nearest-neighbor interactions} \label{subsection:Theory1}

We consider the problem of interacting hard rods of length $a=147$ bp confined to a one-dimensional lattice of length $L$ bp (the length of the DNA segment)
(Fig.~\ref{Fig:NucleosomesSketch}).
Let $u(k)$ be the external potential energy of a particle that occupies positions $k$ through $k + a - 1$ on the DNA (the one-body energy),
and let $\Phi(k,l)$ be the two-body interaction between a pair of nearest-neighbor particles with starting positions $k$ and $l$, respectively. 
Here $u(k)$ describes intrinsic histone-DNA interactions, while $\Phi(k,l)$ accounts for the effects of chromatin structure. 
We assume that the DNA segment is surrounded by impenetrable walls, so that
\[
u(0)=u(L-a+2)=u(L-a+3)=\ldots=u(L)=\infty.
\]
Moreover, particle overlaps are not allowed and the two-body potential is short-range as in the Takahashi hard-rod model \cite{Takahashi:1942},
\[
\Phi(k,l)= \left\{ \begin{array}{cl}
\infty & \text{if }l<k+a,\\
0 & \text{if }l\geq k+2a. \end{array} \right.
\]

The canonical partition function  for a fixed number of particles $N$ is given by
\begin{equation} \label{eq:QN}
	Q_N=\sum_{ i_m+a\leq i_{m+1} } e^{-\beta u(i_1)} e^{-\beta \Phi(i_1,i_2)} e^{-\beta u(i_2)} \ldots e^{-\beta u(i_{N-1})} e^{-\beta \Phi(i_{N-1},i_N)} e^{-\beta u(i_N)},
\end{equation}
where $\beta = 1/k_B T$ is the inverse temperature.

Let us introduce two $l_{\text{max}}\times l_{\text{max}}$ matrices,
where $l_{\text{max}}=L-a+1$ is the rightmost starting position of a particle of length $a$:
\begin{align*}
	\la k \ve w \ve l\ra &=
					\begin{cases} 
							e^{-\beta \Phi(k,l)} &\text{if $l \geq k+a$,} \\
							0 &\text{otherwise}, 
					\end{cases} \\
	\la k \ve e \ve l\ra &=e^{-\beta u(k)} \delta_{k,l}.
\end{align*}
Here $\delta_{k,l}$ is the Kronecker delta symbol, and $\la k \ve M \ve l \ra$ represents the element of matrix $M$ in row $k$ and column $l$ (in Dirac notation).
$\ve l \ra$ is a column vector of dimension $l_{\text{max}}$ with $1$ at position $l$ and $0$ everywhere else, and
$\la k \ve$ is a row vector with $1$ at position $k$.

Defining $\ve J \ra=\sum_{l=1}^{l_{\text{max}}} \ve l \ra$ (a vector with $1$ at every position), we rewrite Eq.~\eqref{eq:QN} as
\[
	Q_N=\begin{cases} 
							\la J\ve (e w)^{N-1} e \ve J\ra &\text{if $N \geq 1$,} \\
							1 &\text{if $N=0$.} 
					\end{cases}
\]

The grand-canonical partition function is then given by
\begin{equation}
	Z = \sum_{N=0}^{N_\text{max}} e^{\beta N \mu} Q_N = 1 + \la J \ve (I-z w)^{-1} z \ve J \ra, \label{Eq:Z}
\end{equation}
where $\mu$ is the chemical potential, $N_\text{max}=\left\lfloor \frac{L}{a}\right\rfloor$ is the maximum number of particles that can fit on $L$~bp,
$I$ is the identity matrix, and $\la k \ve z \ve l\ra = e^{\beta [\mu-u(k)]} \delta_{k,l}$.

The $s$-particle distribution functions are defined as
\[
	n_s(i_1, \ldots, i_s) \equiv \frac{\xi(i_1) \ldots \xi(i_s)}{Z} \frac{\delta^s Z}{\delta \xi(i_1) \ldots \delta \xi(i_s)},
\]
where $\xi(i)=e^{\beta [\mu - u(i)]}$ (see the chapter by Stell in \cite{FL:1964}).
The one-particle distribution function is
\begin{equation} \label{Eq:n}
	n(i) = \frac{1}{Z} \la J \ve (I - z w)^{-1} \ve i \ra \la i \ve z \ve i \ra \la i \ve (I - w z)^{-1} \ve J \ra,
\end{equation}
and the two-particle, nearest-neighbor distribution function is
\begin{equation} \label{Eq:n2}
	\overline{n}_2(i,j) = \frac{1}{Z} \la J \ve (I - z w)^{-1} \ve i \ra \la i \ve z w z \ve j \ra \la j \ve (I- w z)^{-1} \ve J \ra.
\end{equation}

These relations are easy to understand. To find the probability of starting a particle at position $i$ [Eq.~\eqref{Eq:n}], we have to add the statistical weights
of all configurations that contain a particle at that position, and divide the resulting sum by the partition function.
Similarly, to find the probability of having a pair of nearest-neighbor particles with the starting positions $i$ and $j$ [Eq.~\eqref{Eq:n2}], we
need to sum the statistical weights of all configurations that contain that pair of particles. 

Note that for short distances $j-i<2a$, $\overline{n}_2(i,j)$ is identical to the unrestricted two-particle distribution $n_2(i,j)$, because
there is not enough space to put another particle between the two particles at $i$ and $j$. In this paper we
restrict ourselves to $j-i<2a$, since we are interested in short-range interactions
between nearest-neighbor pairs of particles.

In many cases of interest the energetics of the system is unknown but the $s$-particle distributions are available from experiment. Therefore, we wish
to find the unknown energies $u$ and $\Phi$ from $n$ and $\overline{n}_2$ by inverting Eqs.~\eqref{Eq:n} and \eqref{Eq:n2}.
Let us define two matrices: $\la i \ve N \ve j \ra = n(i) \delta_{i,j}$ and $\la i \ve N_2 \ve j \ra = \overline{n}_2(i,j)$.
Using these matrices and Eq.~\eqref{Eq:Z}, we have
\[
\la J \ve (I- N_2 N^{-1}) N \ve J \ra = \frac{1}{Z} \la J \ve (I - z w)^{-1} z \ve J \ra = \frac{Z-1}{Z},
\]
so that
\begin{equation}
 Z = \frac{1}{1 - \la J \ve (I- N_2 N^{-1}) N \ve J \ra}. \label{Eq:Z2}
\end{equation}

After some matrix multiplications we obtain the following two equalities:
\begin{equation}
\la J \ve I - N_2 N^{-1} \ve k \ra = \frac{1}{\la J \ve (I - z w)^{-1} \ve k \ra}, \label{Eq:Inversion1}
\end{equation}
and
\begin{equation}
	\la k \ve I- N^{-1} N_2 \ve J \ra = \frac{1}{\la k \ve (I - w z)^{-1} \ve J \ra}. \label{Eq:Inversion2}
\end{equation}

Substituting Eqs.~\eqref{Eq:Z2}, \eqref{Eq:Inversion1} and \eqref{Eq:Inversion2} into Eqs.~\eqref{Eq:n} and \eqref{Eq:n2}, we derive the exact 
expressions for one-body energies and two-body interactions \cite{Percus:1976,Percus:1989}:
\begin{equation}
-\beta \left[u(k)-\mu\right] = \ln \left( \frac{\la J \ve I - N_2 N^{-1} \ve k \ra \la k \ve N \ve k \ra \la k \ve I- N^{-1} N_2 \ve J \ra}{1 - \la J \ve (I- N_2 N^{-1}) N \ve J \ra}\right), \label{Eq:u}
\end{equation}
\begin{equation}
-\beta \Phi(k,l) = \ln \left( \frac{\la k \ve N^{-1} N_2 N^{-1} \ve l \ra \left[ 1 - \la J \ve (I- N_2 N^{-1}) N \ve J \ra \right]}
	{\la k \ve I- N^{-1} N_2 \ve J \ra \la J \ve I - N_2 N^{-1} \ve l \ra}\right). \label{Eq:Phi}
\end{equation}

If the two-body interactions other than the steric exclusion are neglected, then $\la k \ve w \ve l\ra \to \Theta(l-k-a)$, where $\Theta(l-k-a)$
is the Heaviside step function ($1$ if $l \geq k+a$, and $0$ otherwise). 
In this case we have:
\begin{align*}
\la J \ve (I - z w)^{-1} \ve i \ra &= \la J \ve I + z w + (z w)^2 + \ldots \ve i \ra = Z_{i-1}^f,\\
\la i \ve (I - w z)^{-1} \ve J \ra &= \la i \ve I + w z + (w z)^2 + \ldots \ve J \ra = Z_{i+a}^r,\\
\end{align*}
where $Z_i^f$ and $Z_i^r$ are partial statistical sums which can be efficiently computed by iteration in a forward or reverse direction
\cite{Morozov:2009,Locke:2010}. Note that $Z = Z_1^r = Z_{L-a+1}^f$.
The partial statistical sums account for the contributions from all possible configurations of particles
confined to the boxes $[1,i]$ and $[i,L]$, respectively. It can be shown that \cite{Locke:2010}:

\begin{align*}
Z_i^f &= \prod_{j=1}^{i} \frac{1 - O(j+1) + n(j+1)}{1 - O(j)}, \\
Z_i^r &= \prod_{j=i}^{L} \frac{1 - O(j) + n(j)}{1 - O(j)},
\end{align*}
where $O(i)$ is the particle occupancy of bp $i$ $\left[O(i) = \sum_{j = i-a+1}^{i} n(j)\right]$.

Using Eq.~\eqref{Eq:n}, we reproduce the previous result from~\cite{Locke:2010} which can be employed to find one-body energies
from one-particle distribution in the case of hard-core interactions alone,
\begin{align}
-\beta \left[u^{0}(i) - \mu \right] =& \ln \left[ \frac{n(i)}{1 - O(i) + n(i)}\right] + \ln \left[\prod_{j=i}^{i+a-1} \frac{1 - O(j) + n(j)}{1 - O(j)} \right]. \label{Eq:u0}
\end{align}
\subsection{Predicting two-body interactions from one-particle distribution} \label{subsection:Theory2}

As shown above, there is a one-to-one correspondence between one-body energies $u$ and two-body interactions $\Phi$ on one hand, and
particle distributions  $n$ and $\overline{n}_2$ on the other. Thus, if $n$ and $\overline{n}_2$ are known, $u$ and $\Phi$
can be inferred exactly, and vice versa. However, in many situations the two-particle distribution is not directly available from experiments.
For example, high-throughput nucleosome maps simultaneously report nucleosome positions from many cells, effectively yielding
a probabilistic description of the one-particle distribution $n$. Because of this averaging over single-cell configurations,
information about the pair density profile $\overline{n}_2$ cannot be extracted directly.
Nonetheless, if the two-body interactions are sufficiently strong, the one-particle distribution profile $n$ can be used to obtain information about $\Phi$.

Let us introduce the dimensionless pair distribution
\begin{equation}
g(i,j)=\frac{n_2(i,j)}{n(i) n(j)}. \label{Eq:LinkerAprox}
\end{equation}

Note that $g(i,j) = \overline{n}_2(i,j)/[n(i) n(j)]$ for short distances $j-i<2a$, and that  $g(i,j)=g(j-i)$
in a homogeneous system. We start with a homogeneous system of $N$ hard-rods
which interact through an arbitrary nearest-neighbor potential $\Phi$, and then develop an approximation for the inhomogeneous case.
In a translation-invariant continuous system with nearest-neighbor interactions of arbitrary strength,
$e^{-\beta \Phi(d)}=C e^{\alpha d} g(d)$, where $C$ and $\alpha$ are constants
\cite{Takahashi:1942,Gursey:1950,Salsburg:1953,Fisher:1964}. The result can also be proved for a lattice fluid of hard rods,
as shown below.

\begin{proof}
Consider a system of $N$ particles distributed on a segment of length $L$ bp. We assume that the particles interact with each other through 
short-range nearest-neighbor interactions (which include steric exclusion if the particles have a finite size of $a$ bp), and the total interaction energy is
\[
U(x_1, x_2, \ldots, x_N)=\Phi(x_2 - x_1) + \Phi(x_3 - x_2) + \ldots + \Phi(x_N - x_{N-1}) + U_b,
\]
where $U_b$ is the boundary term which describes interaction between the walls and the first and last particles. 

For simplicity let us assume that the boundary conditions are enforced by two additional particles of the same kind fixed at $x=0$ and 
$x=L$, so that $U_b=\Phi(x_1) + \Phi(L-x_N)$. The exact form of boundary conditions is not essential in the thermodynamic limit.
The canonical partition function of this system of $N$ particles is
\begin{align*}
Q_N(L) &= \sum_{0 \leq x_1 \leq x_2 \leq \ldots \leq x_N \leq L} e^{-\beta \Phi(x_1 - 0)} e^{-\beta \Phi(x_2 - x_1)} \ldots e^{-\beta \Phi(L - x_N)}\\
	&= \sum_{x_N = 0}^L \sum_{x_{N-1} = 0}^{x_N} \ldots \sum_{x_1 = 0}^{x_2} e^{-\beta \Phi(x_1 - 0)} e^{-\beta \Phi(x_2 - x_1)} \ldots e^{-\beta \Phi(L - x_N)}
\end{align*}

Denoting $f(x) \equiv e^{-\beta \Phi(x)}$, we obtain
\[
Q_N(L) = \sum_{x_N = 0}^L \sum_{x_{N-1} = 0}^{x_N} \ldots \sum_{x_1 = 0}^{x_2} f(x_1-0) f(x_2 - x_1) \ldots f(L - x_N).
\]
Note that this represents the convolution of $N+1$ functions $f$,
\[
Q_N(L) = (\underbrace{f * f * \ldots * f}_{N+1 \text{ functions}})(L).
\]

The partition function can be computed using the $z$ transform method. Let $\widetilde{Q}(z)$ be the $z$ transform of $Q_N(L)$,
\[
\widetilde{Q}(z) = \sum_{n=0}^\infty Q_N(n) z^{-n}.
\]

From the convolution theorem we have that
\[
\widetilde{Q}(z) = \left[ \widetilde{F}(z) \right]^{N+1},
\]
where $\widetilde{F}(z)$ is the $z$ transform of $f(n)$,
\[
\widetilde{F}(z) = \sum_{n=0}^\infty f(n) z^{-n}.
\]

The partition function can be recovered using the inverse $z$ transform,
\[
Q_N(L) = \frac{1}{2 \pi i} \oint_\Gamma \left[ \widetilde{F}(z) \right]^{N+1} z^{L-1} \d z.
\]
The contour of integration $\Gamma$ is any simple closed curve enclosing $|z|=R$, $|z|>R$ being the region of convergence.

Let us define $h(z)=(N+1) \ln \widetilde{F}(z) + (L-1) \ln z$. With this notation
\[
Q_N(L)=\frac{1}{2 \pi i} \oint_\Gamma e^{h(z)} \d z.
\]

This integral can be computed by the saddle point method \cite{Fisher:1964}. Expanding $h(z)$ around the saddle point $z_0$, we obtain
\[
Q_N(L)\approx e^{h(z_0)} \frac{1}{2 \pi i} \int e^{\frac{1}{2} h''(z_0)(z-z_0)^2} \d z.
\]
Integration along the path of steepest descent yields a contribution from the Gaussian integral of order $O\left( [h''(z_0)]^{-1/2} \right)=O\left( N^{-1/2} \right)$.
Since we need $\ln Q_N(L)$ in order to compute the macroscopic quantities, and in the thermodynamic limit the terms of order $O(\ln N)$ are not important, we can approximate the partition function as
\begin{equation}
Q_N(L) \approx e^{h(z_0)} \approx z_0^{L} \left[ \widetilde{F}(z_0) \right]^{N}, \label{Eq:PartF}
\end{equation}
where $z_0$ is the saddle point, given by
\begin{equation}
\left.\frac{\d h}{\d z}\right|_{z=z_0} \approx \frac{L}{z_0} + N \frac{\widetilde{F}'(z_0)}{\widetilde{F}(z_0)} = 0.\label{Eq:SaddlePoint}
\end{equation}

We can compute the chemical potential for the interacting hard rods by taking the derivative of the free energy $F$ with respect to the number of particles in the system
\begin{equation}
\mu = \frac{\partial F}{\partial N} = -k_B T \frac{\partial \ln Q_N}{\partial N} = -k_B T \ln \widetilde{F}(z_0). \label{Eq:ChemPot}
\end{equation}

The pressure of the gas is given by the derivative of the free energy with respect to the length of the system. Let's denote the length of a base pair by $b$, such that the real length of the system is $L b$. We obtain
\begin{equation}
p = -\frac{1}{b}\frac{\partial F}{\partial L} = \frac{k_B T}{b} \frac{\partial \ln Q_N(L)}{\partial L} = \frac{k_B T}{b} \ln z_0, \label{Eq:Pressure}
\end{equation}
and from Eq.~\eqref{Eq:PartF} we obtain
\[
Q_N(L)=\left( e^{\beta p b} \right)^L \left[\widetilde{F} \left(e^{\beta p b}\right)\right]^N.
\]

We use this result to compute the conditional probability of finding an adjacent particle
at a distance $d$ from the center of a fixed particle \cite{Gursey:1950,Tonks:1936}:
\begin{align*}
P(d)\equiv&~\text{Prob}(x_N=L-d|x_{N+1}=L)\\
=&\frac{1}{Q_N} \sum_{x_{N-1} = 0}^{x_N} \ldots \sum_{x_1 = 0}^{x_2} f(x_1) \ldots f\boldsymbol{(}(L-d) - x_{N-1}\boldsymbol{)} f\boldsymbol{(}L-(L-d)\boldsymbol{)}\\
=&f(d)\frac{Q_{N-1}(L-d)}{Q_N(L)}\\
=&e^{-\beta \Phi(d)} \frac{(e^{\beta p b})^{-d}}{\widetilde{F}(e^{\beta p b})}.
\end{align*}

For $d<2a$ the pair distribution becomes
\[
g(d)=\frac{\overline{n}_2(i, i+d)}{n(i)n(i+d)}=\frac{\rho P(d)}{\rho^2} \propto e^{-\beta p b d} e^{-\beta \Phi(d)}.
\]

Thus in the homogeneous system the interaction between the particles has the form
\begin{equation} \label{eq:ePhi}
e^{-\beta \Phi(d)} = C e^{\alpha d} g(d),
\end{equation}
where $\alpha=\beta p b$ and $C$ is a normalization constant. \qed
\end{proof}

In a more general case, the external potential breaks translational invariance, making $g$ dependent on the absolute position of the first particle.
However, if the two-body interaction $\Phi$ is translationally invariant, a good approximation is provided by
replacing $g$ with $P_{\text{linker}} (\Delta) = \la g(i,i+a+\Delta) \ra_i$ (averaged over all initial positions $i$) in Eq.~\eqref{eq:ePhi} \cite{Chereji:2010},
\begin{equation} \label{Eq:Phi:est}
-\beta \Phi (i,j) \approx \ln \left[P_{\text{linker}}\boldsymbol{(}j-(i+a)\boldsymbol{)}\right] + \alpha (j-i) + \ln C 
\end{equation}
The constants $C$ and $\alpha$ are uniquely determined by the asymptotic condition $\lim_{(j-i) \to \infty} \Phi(i,j)= 0$.
Eq. \eqref{Eq:Phi:est} provides an ansatz for reconstructing $\Phi$ from $P_{\text{linker}}(\Delta)=\la \overline{n}_2(i,i+a+\Delta)/[n(i)n(i+\Delta)] \ra_i$.

\begin{figure}[t]
	\begin{center}
	\includegraphics[width=\textwidth]{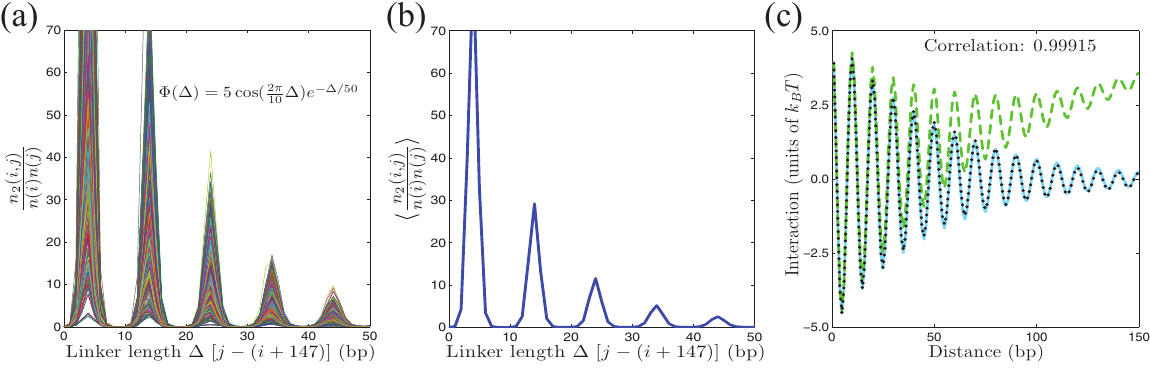}   
	\end{center} 
	\caption{(Color online) (a) $g(i,j)={n_2(i,j)}/{n(i)n(j)}$ is plotted for a representative subset of all initial positions $i$
in a $10^4$ bp DNA segment. The one-body energies are randomly sampled from a Gaussian distribution with a mean of $2.5~k_B T$ and
a standard deviation of $0.2~k_B T$, and $9$ potential wells of depth $5~k_B T$ are added at $1, 2, \dots ,9$ $\times 10^3$ bp to model a strongly inhomogeneous
system. $n_2(i,j)$ and $n(i)$ are computed from one- and two-body energies using Eqs.~\eqref{Eq:n} and \eqref{Eq:n2}.
(b) $P_\textrm{linker} (\Delta)$, obtained by averaging $g(i,j)$ over all initial positions $i$.
Note that $\Delta=j-(i+147)$ represents the linker length between the two nucleosomes with starting positions $i$ and $j$, respectively.
(c) Exact (solid blue line) and predicted (dotted black line) two-body interactions. The predicted interaction was computed from the $-\ln(P_{\text{linker}})$ curve 
(dashed green line) using Eq.~\eqref{Eq:Phi:est}.}
\label{Fig:ModelApprox}
\end{figure}

Fig.~\ref{Fig:ModelApprox} shows a numerical test of this ansatz on a $10$ kbp DNA segment. We construct a random one-body energy landscape
and simulate strong inhomogeneity by positioning $9$ potential wells with depth of $5 k_B T$ at $1, 2, \dots ,9$ kbp on the landscape. 
The model interaction between a pair of particles separated by a linker of length $\Delta$ is $\Phi(\Delta) = 5 \cos \left( \frac{2 \pi}{10} \Delta \right) e^{-\Delta/50}$ (in units of $k_B T$). We use the one-body energies and the two-body potential as inputs to
Eqs.~\eqref{Eq:n} and \eqref{Eq:n2}, computing the dimensionless pair distribution function. The pair distribution
varies significantly from bp to bp [Fig.~\ref{Fig:ModelApprox}(a)],
as can be expected in a system with one- and two-body energies of comparable magnitude. Following our prescription, we compute $P_{\text{linker}}$
by averaging over all the curves in Fig.~\ref{Fig:ModelApprox}(a) [Fig.~\ref{Fig:ModelApprox}(b)], and employ Eq.~\eqref{Eq:Phi:est} to infer $\Phi$
[Fig.~\ref{Fig:ModelApprox}(c)]. The correlation coefficient between predicted and exact two-body interactions is greater than $0.999$.

If a direct measurement of the pair distribution $\overline{n}_2$ is not available, 
$P_{\text{linker}}$ needs to be estimated empirically from the $n$ profile.
Each nucleosome positioning data set consists of the histogram of the number of nucleosomes starting at each genomic bp $i$.
We preprocess these data by removing all counts of height $1$ from the histogram and smoothing
the remaining counts with a $\sigma=2$ Gaussian kernel.
Next, we compute $n(i)$ by rescaling the smoothed profile so that the maximum occupancy for each chromosome is $1$.
Finally, we identify all local maxima on the $n$ profile and assume that they mark prevalent nucleosome positions.
Specifically, for each maximum at bp $i$ we find subsequent maxima at positions $i+146<j_1<j_2<j_3<\dots$ in the $50$ bp window.
To each pair of maxima $(i,j_1),~(i,j_2), \dots$ we assign the probability that they represent neighboring nucleosomes: $n(i) n(j_1),~n(i) [1-n(j_1)] n(j_2)$, and so on.
We sum the probabilities over all initial positions $i$ and normalize, producing an empirical estimate of $P_{\text{linker}}$.

\subsection{Comparison between lattice and continuous one-dimensional fluids}
The formalism presented in the previous subsection can be used to obtain the equation of state and chemical potential for any generic interaction $\Phi(x)$.
Let us consider two simple cases: the ideal gas and the Tonks lattice gas, which is characterized by the hard-core interaction
\[
\Phi(x)=
\left\{
\begin{array}{ll}
\infty &\text{ if } x<a,\\
0 &\text{ if } x \geq a.
\end{array} \right.
\]

For the ideal gas, the $z$ transform of $e^{-\beta \Phi}$ and the saddle point $z_0$ [obtained from Eq.~\eqref{Eq:SaddlePoint}] are given by:
\begin{align*}
\widetilde{F}(z) &= \frac{z}{z-1},\\
z_0 &= \frac{L+N}{N}.
\end{align*}

Using these expressions, we can compute the logarithm of the partition function [Eq.~\eqref{Eq:PartF}]
\[
\ln Q_N(L) = L \ln \left(\frac{L+N}{L}\right) + N \ln \left(\frac{L+N}{N}\right) + O(\ln N),
\]
which gives the pressure and the chemical potential for the ideal lattice gas:
\begin{align}
\beta p_\text{l}^\text{id} &= \frac{1}{b} \ln \left( 1+\frac{N}{L}\right),\label{Eq:PressureLatticeIdeal}\\
\beta \mu_\text{l}^\text{id} &= \ln \left(\frac{N}{L+N}\right).\label{Eq:MuLatticeIdeal}
\end{align}

In the case of the Tonks lattice gas,
\begin{align*}
\widetilde{F}(z) &= \frac{z^{1-a}}{z-1},\\
z_0 &= \frac{L-N a+N}{L-N a}.
\end{align*}

The pressure and the chemical potential are then given by:
\begin{align}
\beta p_\text{l}^\text{T} &= \frac{1}{b} \ln \left( 1+\frac{N}{L-N a}\right),\label{Eq:PressureLatticeTonks}\\
\beta \mu_\text{l}^\text{T} &= a \ln \left(\frac{L-N a+N}{L-N a}\right)+\ln \left(\frac{N}{L-N a+N}\right). \label{Eq:MuLatticeTonks}
\end{align}

It is useful to compare these results with the corresponding results for continuous one-dimensional gases.
Denoting the physical length of the particles by $\mathcal{A}$, the length of the box by $\mathcal{L}$, and
the Laplace transform of $e^{-\beta \Phi}$ by
\[
\varphi(s) = \int_0^\infty s^{-s x} e^{-\beta \Phi(x)} \d x,
\]
we obtain the canonical partition function as the inverse Laplace transform
\[
Q_\text{cont}(N, \mathcal{L}, T)=\frac{1}{\lambda(T)^N} \frac{1}{2 \pi i} \int_{c-i\infty}^{c+i\infty}e^{s \mathcal{L}} \left[ \varphi(s)\right]^N \d s \approx e^{s_0 \mathcal{L}} \left[ \frac{\varphi(s_0)}{\lambda(T)} \right]^N,
\]
where $\lambda(T)={h}/{\sqrt{2 \pi m k_B T}}$ is the thermal de Broglie wavelength, and $s_0$ is the saddle point given by the equation
\[
\mathcal{L}+N\frac{\varphi'(s_0)}{\varphi(s_0)}=0.
\]

Using the previous two equations, we find for the ideal gas $\big[\varphi(s) = \frac{1}{s}$, and $s_0=\frac{N}{\mathcal{L}}\big]$:
\begin{align}
\beta p_\text{c}^\text{id} &= \frac{\partial \ln Q_{\text{cont}}}{\partial \mathcal{L}} = \frac{N}{\mathcal{L}},\label{Eq:PressureContIdeal}\\
\beta \mu_\text{c}^\text{id} &= - \frac{\partial \ln Q_{\text{cont}}}{\partial N} = \ln \frac{N \lambda(T)}{\mathcal{L}}.\label{Eq:MuContIdeal}
\end{align}

Similarly, for the Tonks gas we obtain $\Big[\varphi(s) = \frac{e^{-\mathcal{A} s}}{s}$, and $s_0=\frac{N}{\mathcal{L}-N \mathcal{A}}\Big]$:
\begin{align}
\beta p_\text{c}^\text{T} &= \frac{N}{\mathcal{L}-N \mathcal{A}},\label{Eq:PressureContTonks}\\
\beta \mu_\text{c}^\text{T} &= \frac{N \mathcal{A}}{\mathcal{L}-N \mathcal{A}}+\ln\left[\frac{N \lambda(T)}{\mathcal{L} - N \mathcal{A}}\right].\label{Eq:MuContTonks} 
\end{align}

To compare continuous and discrete results, we let the lattice constant $b$ approach $0$, while keeping the particle size ($\mathcal{A} = a b$) and the box size ($\mathcal{L} = L b$)
finite. We obtain:
\begin{align*}
\lim_{b \to 0} \beta p_\text{l}^\text{id}&
= \lim_{b \to 0} \frac{1}{b} \ln \left( 1+\frac{N b}{\mathcal{L}}\right) = \beta p_\text{c}^\text{id},\\
\lim_{b \to 0} \beta p_\text{l}^\text{T}&=\lim_{b \to 0} \frac{1}{b} \ln \left( 1+\frac{N b}{\mathcal{L} - N \mathcal{A}}\right) = \beta p_\text{c}^\text{T}.
\end{align*}

Similarly, the chemical potentials for the ideal and Tonks lattice gases become asymptotically, as $b \to 0$:
\begin{align*}
\beta \mu_\text{l}^\text{id} &\sim \ln \frac{N b}{\mathcal{L}},\\
\beta \mu_\text{l}^\text{T} &\sim \frac{N \mathcal{A}}{\mathcal{L}-N \mathcal{A}}+\ln\left(\frac{N b}{\mathcal{L} - N \mathcal{A}}\right).
\end{align*}
These expressions are identical to the chemical potentials of the corresponding continuous gases [Eqs.~\eqref{Eq:MuContIdeal} and \eqref{Eq:MuContTonks}],
with the length scale $\lambda(T)$ replaced by the typical length scale of the lattice, $b$.

\subsection{Sequence-specific energy of nucleosome formation} \label{subsection:Theory3}

We can extract a sequence-specific component of the one-body energy by using Eqs.~\eqref{Eq:u} or \eqref{Eq:u0} to compute $u - \mu$, estimating the chemical
potential $\mu$, and fitting the one-body energy $u$ to a linear model
which assigns energies to nucleotide words found within the $a = 147$ bp nucleosomal site.
Assuming that the system is nearly homogeneous,
we use Eqs.~\eqref{Eq:ChemPot} or \eqref{Eq:MuLatticeTonks} to obtain the chemical potential of the lattice gas.
After eliminating $\mu$, we fit a linear model to one-body energies $u$.
It was established in Ref.~\cite{Locke:2010}
that position-independent models in which the energy of the nucleotide word does not depend on its exact location within the nucleosome
can be used to describe genome-wide nucleosome occupancies.
Furthermore, an $N=2$ position-independent model with just $13$ fitting parameters performed as well as $N > 2$ models
[here, $N$ denotes the longest word (in bp) included into the model].

If both monomers and dimers contribute to the total one-body energy, the sequence-specific binding energy of a $147$ bp-long nucleosomal site is given by
\begin{equation} \label{Eq:EnergyN2}
u^S = \sum_{\alpha} m_{\alpha} \epsilon_{\alpha} + \sum_{\alpha, \beta} m_{\alpha \beta} \epsilon_{\alpha \beta}
+ \epsilon_{0},
\end{equation}
where $m_\alpha$ is the number of nucleotides of type $\alpha\in\{A,C,G,T\}$, $\epsilon_\alpha$ is the energy of the nucleotide $\alpha$, and $\epsilon_0$ 
is the overall sequence-independent offset.
Similarly, $m_{\alpha \beta}$ is the number of dinucleotides of type $\alpha \beta$, and $\epsilon_{\alpha \beta}$ is the corresponding energy. 
In Ref.~\cite{Locke:2010}, word energies were constrained by
$\sum_{\alpha} \epsilon_{\alpha} = \sum_{\alpha} \epsilon_{\alpha \beta} = \sum_{\beta} \epsilon_{\alpha \beta} = 0$,
yielding a $13$-parameter model.
Here we develop an alternative approach which does not impose any additional constraints beyond those caused by the fact that the number of mono- and dinucleotides
in the $147$ bp-long site is fixed.

We can express the nucleosome energies as $\mathbf{u}=\mathbf{M}\mathbf{x}$, or equivalently
\[
\left(
\begin{array}{c}
	u^S(1)\\
	u^S(2)\\
	\vdots\\
	u^S(l_{max})
\end{array}
\right)
=\left(
\begin{array}{cccc}
	m_{1,1} & \cdots & m_{1,20} & 1\\
	m_{2,1} & \cdots & m_{2,20} & 1\\
 	\vdots &        & \vdots &\vdots\\
	m_{l_{max},1} & \cdots & m_{l_{max},20} & 1
\end{array}
\right)
\left(
\begin{array}{c}
	x_1\\
	\vdots\\
	x_{21}
\end{array}
\right),
\]
where $u^S(i)$ is the sequence-specific energy of the nucleosome that covers the DNA sequence between bps $i$ and $i+a-1$.
$m_{i,1}, \ldots, m_{i,4}$ give the number of $A, C, G$ and $T$ nucleotides found in that sequence,
and $m_{i,5}, \ldots, m_{i,20}$ give the number of dinucleotides $AA, AC, \ldots, TG, 
TT$. $l_{max} = L-a+1$ is the maximum starting position for a nucleosome, and the set of parameters $x_1,\ldots x_{21}$ represents the $21$ energies from
Eq.~\eqref{Eq:EnergyN2}: $x_1=\epsilon_A$, $x_2=\epsilon_C$, $\ldots$, $x_{20}=\epsilon_{TT}$, $x_{21}=\epsilon_0$.

Note that for any DNA sequence of length $147$~bp,
\begin{align*}
m_{i,1}+m_{i,2}+m_{i,3}+m_{i,4}&=147~m_{i,21},\\
m_{i,5}+m_{i,6}+\ldots+m_{i,20}&=146~m_{i,21}.
\end{align*}
This means that the rank of $\mathbf{M}$ is $19$. For any linear operator $(\mathbf{M})$,
the dimension of its domain ($21$ in our case) is equal to the sum of the dimensions of the image $\operatorname{im}(\mathbf{M})$ in the energy space ($19$)
and the kernel $\ker(\mathbf{M})$ in the parameter space ($2$).

It is easy to check that the vectors
\[
\mathbf{x}^*=
\left(
\begin{array}{c}
	1\\
	1\\
	1\\
	1\\
	0\\
	\vdots\\
	0\\
	-147
\end{array}
\right)\!\!, 
~\mathbf{x}^{**}=
\left(
\begin{array}{c}
	0\\
	0\\
	0\\
	0\\
	1\\
	\vdots\\
	1\\
	-146
\end{array}
\right)
\]
belong to $\ker(\mathbf{M})$, so that $\ker(\mathbf{M}) = \operatorname{span} (\mathbf{x}^*, \mathbf{x}^{**})$.

Any vector of parameters $\mathbf{x}$ can be uniquely decomposed as $\mathbf{x}=\mathbf{x}_K+\mathbf{x}^\perp$, with $\mathbf{x}_K\in \ker(\mathbf{M})$
and $\mathbf{x}^\perp \in \ker(\mathbf{M})^\perp$, the subspace which is orthogonal to $\ker(\mathbf{M})$.
Specifically, $\mathbf{x}^\perp = \mathbf{x} - (\mathbf{x}^*, \mathbf{x})  \mathbf{x}^* - (\mathbf{x}^{**}, \mathbf{x})  \mathbf{x}^{**}$.
The component $\mathbf{x}_K$ does not contribute to the energy of the sequence because $\mathbf{M} \mathbf{x}_K$
is the null vector.
The components of $\mathbf{x}^\perp$ satisfy the following relations:
\begin{align*}
&(\mathbf{x}^\perp, \mathbf{x}^*)=0 \Rightarrow \sum_{i=1}^4 x^\perp_i - 147~x^\perp_{21}=0,\\
&(\mathbf{x}^\perp, \mathbf{x}^{**})=0 \Rightarrow \sum_{i=5}^{20} x^\perp_i - 146~x^\perp_{21}=0.\\
\end{align*}

Thus $\mathbf{x}^\perp$ has only $19$ independent parameters, and
\begin{align*}
x^\perp_{21}&=\frac{1}{147} \sum_{i=1}^4 x^\perp_i,\\
x^\perp_{20}&=146~x^\perp_{21} - \sum_{i=5}^{19} x^\perp_i=\frac{146}{147} \sum_{i=1}^4 x^\perp_i - \sum_{i=5}^{19} x^\perp_i
\end{align*}
\begin{equation} \label{Eq:xperp}
\Rightarrow \mathbf{x}^\perp=
\left(
\begin{array}{c}
	x^\perp_1\\
	\vdots\\
	x^\perp_{19}\\
	\frac{146}{147} \sum_{i=1}^4 x^\perp_i - \sum_{i=5}^{19} x^\perp_i\\
	\frac{1}{147} \sum_{i=1}^4 x^\perp_i
\end{array}
\right)
\end{equation}

In order to compare two different sets of $21$ energies (e.g. fit on different genomes), we need to eliminate the components of these two vectors included in $\ker(\mathbf{M})$.
The components from $\ker(\mathbf{M})^\perp$ will have $19$ independent parameters and $2$ redundant ones [Eq.~\eqref{Eq:xperp}].
The projection of the energy vector on the $\operatorname{im}(\mathbf{M})$ hyperplane is unique, and there is a one-to-one correspondence between 
$\operatorname{im}(\mathbf{M})$ and the parameter subspace which is orthogonal to the kernel, $\ker(\mathbf{M})^\perp$. In this way, every set of fitted energies uniquely 
determines a set of parameters ($\mathbf{x}^\perp$) and a sequence-specific energy ($\mathbf{u}=\mathbf{M} \mathbf{x}^\perp$).

For the $N=1$ model, $\ker(\mathbf{M})$ is spanned by a single vector
\[
\mathbf{x}^*=
\left(
\begin{array}{c}
	1\\
	1\\
	1\\
	1\\
	-147
\end{array}
\right),
\]
and $\mathbf{x}^\perp$ has $4$ relevant parameters and a redundant one
\[
\mathbf{x}^\perp=
\left(
\begin{array}{c}
	x^\perp_1\\	
	x^\perp_2\\
	x^\perp_3\\
	x^\perp_4\\
	\frac{1}{147} \sum_{i=1}^4 x^\perp_i
\end{array}
\right).
\]

Similarly, the $N=3$ model has $85$ ($4+16+64+1$) fitting parameters and there are six independent constraints on the columns of $\mathbf{M}$, so that the rank 
of the operator $\mathbf{M}$ is $79$. The kernel of $\mathbf{M}$ is spanned by six vectors, and the 
parameter subspace orthogonal to the kernel (which gives the sequence energy) is $79$-dimensional. 
For the $N=4$ and the $N=5$ models, the total number of parameters is $341$ and $1365$, and the number of 
independent parameters is $319$ and $1279$, respectively.

When the $N=2$, $21$-parameter model described above is trained on the energies predicted by applying Eq.~\eqref{Eq:u0} to
a large-scale map of nucleosomes reconstituted \textit{in vitro}
on yeast genomic DNA \cite{Zhang:2009}, it captures the same sequence determinants as our previously used $13$-parameter model which employs
additional constraints \cite{Locke:2010} ($r=0.9967$ between the two sequence-specific energy profiles). However, the two approaches are not equivalent,
since the $21$-parameter model utilizes the maximum possible number of independent fitting parameters.
\section{Results}

\subsection{Reconstructing nucleosome energetics in a model system} \label{subsection:Results1}

In the absence of nearest-neighbor interactions induced by chromatin structure, nucleosome formation \textit{in vitro} is fully controlled by
DNA sequence and steric exclusion. In this case efficient procedures are available for reconstructing nucleosome positions from
formation energies \cite{Morozov:2009,Segal:2006} and for inferring nucleosome energetics from experimentally available probability and
occupancy profiles [Eq.~\eqref{Eq:u0}] \cite{Locke:2010}. However, this simple approach may lead to errors if the two-body interactions
are in fact present in the system. Furthermore, many factors other than DNA sequence can affect nucleosome positioning \textit{in vivo}, including
chromatin remodeling enzymes, non-histone DNA-binding factors, and components of transcriptional machinery \cite{Workman:2006,Barrera:2006,Schnitzler:2008}.
These influences are expected to create potential barriers which prevent nucleosomes
from forming in certain regions, and potential wells which localize nucleosomes through favorable contacts between histones and other proteins.
These effects will be lost if a purely sequence-specific model is fit to the nucleosome positioning data.

\begin{figure}[t]
  \begin{center}
  \includegraphics[height=5.1in]{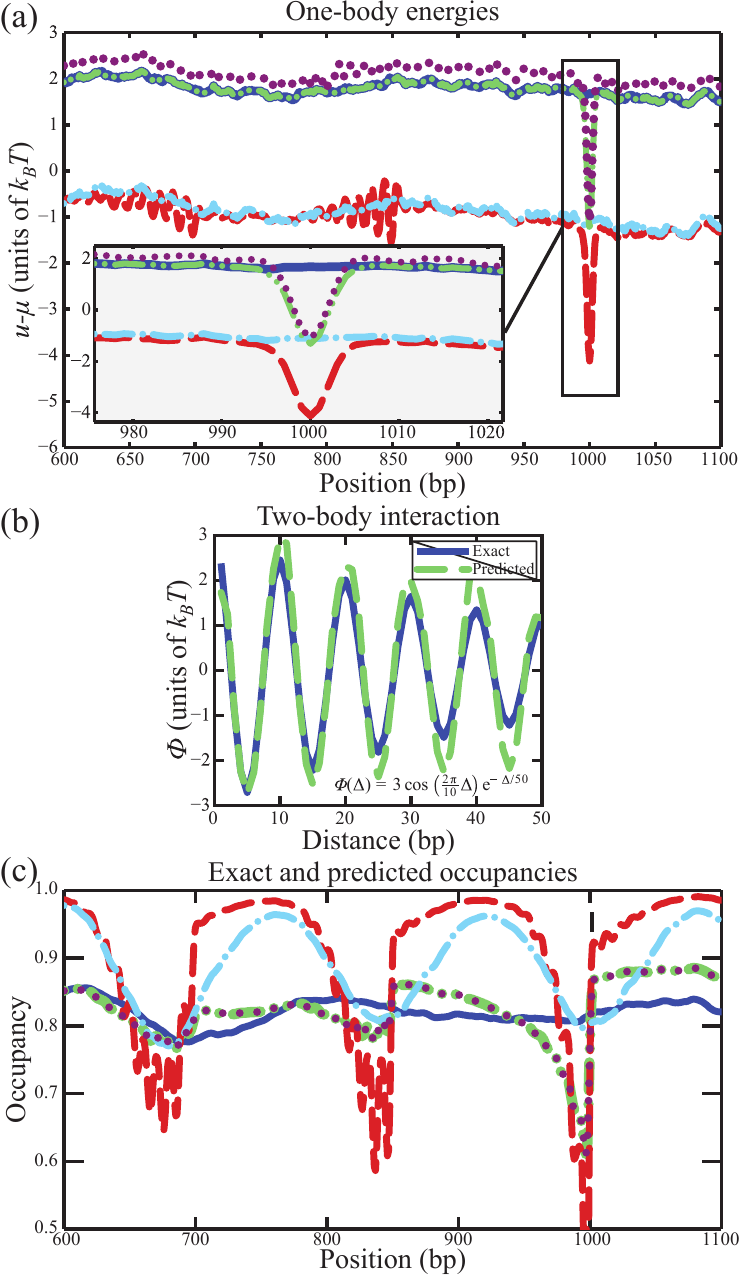}		
  \end{center}
  \caption{(Color online) (a) One-body energies. Sequence-dependent energy given by the $21$-parameter model [Eq.~\eqref{Eq:EnergyN2}] (blue solid line),
total energy given by the sum of the sequence-specific energies and $5$ potential wells with $3~k_B T$ depth at $1, 3, 5, 7$, and $9$ $\times 10^3$ bp
designed to mimic the \textit{in vivo} effects (green dash-dot-dot line).
Energy predicted with a model that neglects two-body interactions [Eq.~\eqref{Eq:u0}] (red dashed line),
energy predicted by fitting the $21$-parameter model to the energies from Eq.~\eqref{Eq:u0} (light blue dash-dot line),
a numerical solution of the full model which takes $\Phi$ into account (maroon dotted line).
Inset: zoom-in on the region with the potential well.
(b) Exact two-body interaction $\Phi$ (blue solid line) and predicted interaction [Eq.~\eqref{Eq:Phi:est}] (green dashed line).
(c) Nucleosome occupancies. Occupancy generated by the exact sequence-specific one-body energy and the exact interaction (blue solid line),
occupancy corresponding to the combined exact one-body energy (sequence-specific component and potential wells) and the exact interaction (green dash-dot-dot line).
Predicted occupancy generated by the one-body energy from Eq.~\eqref{Eq:u0} and predicted $\Phi$ (red dashed line),
occupancy generated using predicted sequence-dependent one-body energy [Eq.~\eqref{Eq:EnergyN2}] and predicted $\Phi$ (light blue dash-dot line),
occupancy predicted using numerically computed one-body energies from the full model and predicted $\Phi$ (maroon dotted line).
}
\label{Fig:Model}
\end{figure}

We use a simple model system to illustrate the errors caused by neglecting higher-order chromatin structure and \textit{in vivo} potentials (Fig.~\ref{Fig:Model}).
We have generated a random $10^4$~bp DNA fragment and computed one-body sequence-dependent energies using the $21$-parameter $N=2$ position-independent
model (section~\ref{subsection:Theory3}). The sequence-specific word energies for the model were randomly sampled from
a $[-0.02~k_B T, 0.02~k_B T]$ uniform distribution.
Fig.~\ref{Fig:Model}(a) shows sequence-dependent nucleosome energies in a representative
$500$~bp window (blue solid line). The window also includes one of the $3~k_B T$ wells placed every 2000 bp throughout the sequence to model
\textit{in vivo} effects; the total one-body energy is shown as a green dash-dot-dot line.

The total energy can be used together with the two-body interaction shown in Fig.~\ref{Fig:Model}(b) (blue solid line)
to construct the exact one-body density profile $n(i)$ and the corresponding nucleosome occupancy for the DNA segment [Fig.~\ref{Fig:Model}(c), green dash-dot-dot line].
If we now use Eq.~\eqref{Eq:u0} which neglects two-body interactions to reconstruct one-body energies from $n(i)$,
the predicted energy profile captures the potential wells and the sequence-specific component
but also displays spurious $10$~bp oscillations caused by the ``leakage'' of the two-body potential $\Phi$ into one-body energetics
[Fig.~\ref{Fig:Model}(a), red dashed line]. In addition, the whole landscape is shifted downward because favorable two-body interactions are missing from the model.
The ``leaked'' oscillations and the \textit{in vivo} wells have no relation to sequence and can be removed by fitting either the $13$- or the $21$-parameter model to the 
prediction [Fig.~\ref{Fig:Model}(a), light blue dash-dot line].
The two predicted energy profiles are highly correlated with each other ($r=0.9993$) and with the exact profile ($r=0.9913$ for the 13-parameter model,
$r=0.9915$ for the 21-parameter model), indicating that the sequence-specific component can be extracted even if the two-body interactions are not handled correctly.

Predicting occupancies from the energy profiles constructed under the $\Phi = 0$ assumption causes discrepancies
with the exact result, shown in Fig.~\ref{Fig:Model}(c) as a green dash-dot-dot line. For example, using the one-body energies predicted
with Eq.~\eqref{Eq:u0} [Fig.~\ref{Fig:Model}(a), red dashed line] and the two-body potential predicted
with Eq.~\eqref{Eq:Phi:est} [Fig.~\ref{Fig:Model}(b), green dashed line] gives an occupancy profile
with higher average occupancy, sharp peaks, and enhanced $10$~bp oscillations compared to the exact
landscape [Fig.~\ref{Fig:Model}(c), red dashed line]. This is not unexpected because the two-body potential is both imprinted in the one-body
profile and included explicitly. In contrast, if $\Phi$ is neglected at this stage as well, the exact occupancy can be restored
from $u^0$ [Eq.~\eqref{Eq:u0} and its inverse do not entail any information loss], but the origin of various contributions remains
unclear as they are all lumped into the one-body landscape.

When the $21$-parameter model is fit to the $\Phi = 0$ profile 
[Fig.~\ref{Fig:Model}(a), light blue dash-dot line] and combined with the predicted $\Phi$ [Fig.~\ref{Fig:Model}(b), green dashed line], the occupancy is off
since \textit{in vivo} potential wells cannot be captured by this model [Fig.~\ref{Fig:Model}(c), light blue dash-dot line]. 
Nucleosomes are not strongly localized if the \textit{in vivo} wells and barriers
are absent [Fig.~\ref{Fig:Model}(c), blue solid line], consistent with the relatively smooth \textit{in vitro} occupancy profiles \cite{Kaplan:2009,Zhang:2009}.
Note that the two occupancy profiles will coincide if the mean of the predicted one-body energies is set to the correct value, eliminating
the spurious offset caused by $\Phi$ [Fig.~\ref{Fig:Model}(a), cf. blue solid and light blue dash-dot lines].

In order to reconstruct the occupancy correctly and avoid mixing one-body and two-body contributions, we need to turn to the full theory
developed in section~\ref{subsection:Theory1}. Inserting predicted $\Phi$ [Fig.~\ref{Fig:Model}(b), green dashed line] into Eq.~\eqref{Eq:n}
and using the exact $n(i)$ profile, we obtain a system of nonlinear equations which can be solved numerically,
yielding energies that are very close to the exact result [Fig.~\ref{Fig:Model}(a), maroon dotted line]. These energies and the predicted
$\Phi$ can be used to reconstruct the occupancy profile which is nearly exact
[Fig.~\ref{Fig:Model}(c), cf. green dash-dot-dot and maroon dotted lines]. Thus we have succeeded in separating one- and two-body effects, and in splitting
off the sequence-dependent part in the former. However, the full procedure is computationally intensive and becomes inefficient
if the DNA is longer than $10^4$ bp (longer segments may be split into manageable pieces and handled separately). 

\begin{figure}[t]
  \begin{center}
  \includegraphics[height=6.3in]{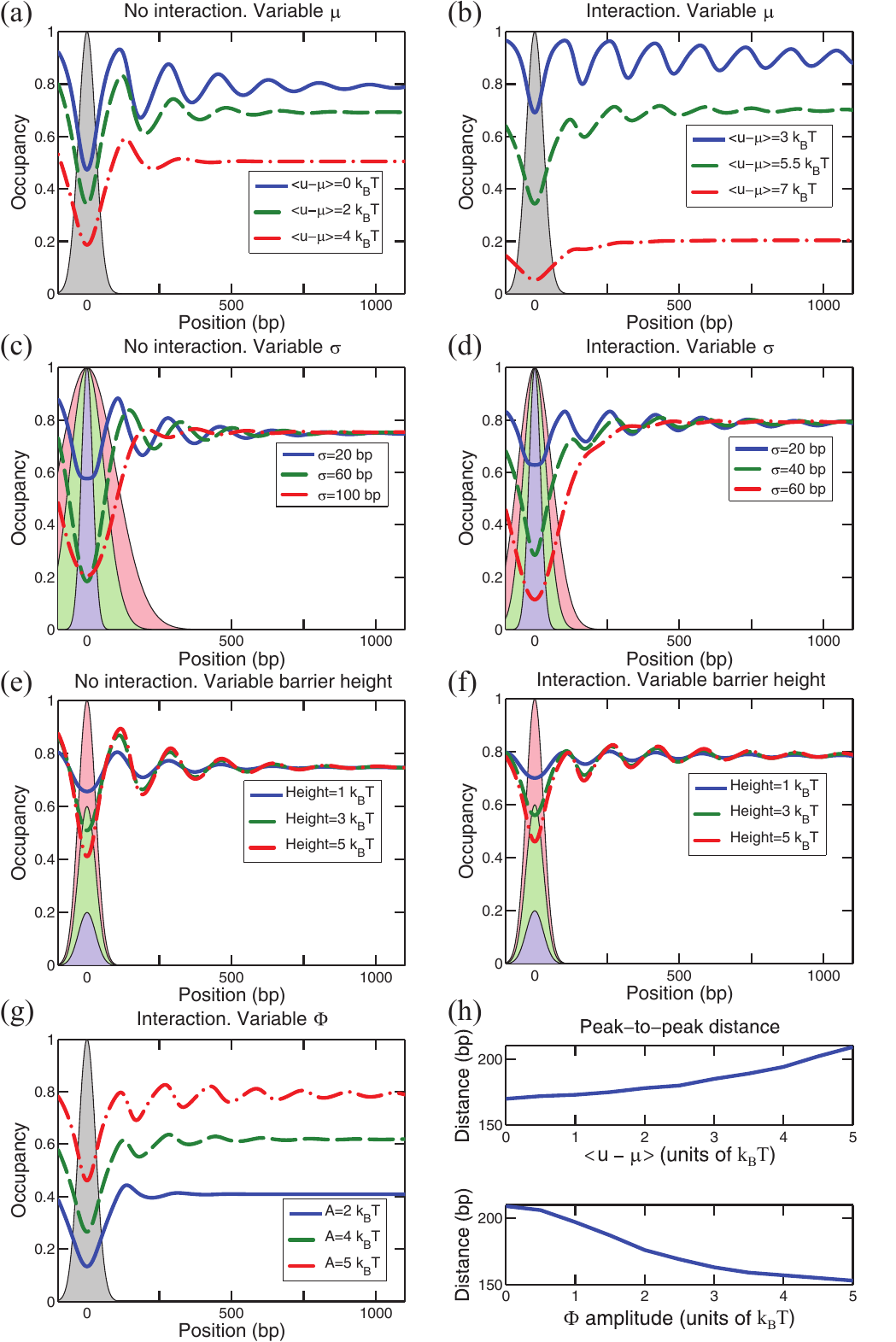}    
  \end{center}
  \caption{(Color online) Symmetric Gaussian barrier. Occupancy profiles for the following scenarios: variable chemical potential $\mu$
(a) and (b), variable barrier width (c) and (d), and variable barrier height (e) and (f). Unless otherwise specified in the legend, the barrier heights are $5~k_B T$,
$\sigma=30$ bp, and $\la u - \mu \ra=5~k_B T$ [in panel (c) $\la u - \mu \ra=1~k_B T$].
Panels (b), (d) and (f) have a two-body interaction
$\Phi(\Delta)=A \cos \left( 2\pi \Delta/10\right) \exp(-\Delta/50)$, with $A=5~k_BT$.
(g) Occupancy profiles for variable interaction strength $A$.
(h) Variation of the typical distance between neighboring nucleosomes as $\mu$ or $A$ is varied. Upper panel: $\Phi = 0$, lower panel: $\langle u \rangle - \mu = 5~k_BT$.
}
\label{Fig:GaussBarrier}
\end{figure}
\subsection{Nucleosome localization by potential barriers and wells} \label{subsection:Results2}

Nucleosomes in the vicinity of potential barriers and wells can be localized by steric exclusion alone \cite{Kornberg:1988}.
This mechanism is thought to contribute to prominent nucleosome occupancy peaks in genic regions observed \textit{in vivo}
but not \textit{in vitro} \cite{Kaplan:2009,Zhang:2009,Yuan:2005,Mavrich:2008a,Mavrich:2008b,Zawadzki:2009}.
In order to understand the nature and the extent of \textit{in vivo} nucleosome localization, we need to
study nucleosome occupancy patterns created by placing a single potential barrier or potential well onto an otherwise flat one-body energy landscape.

\begin{figure}[t]
  \begin{center}
  \includegraphics[height=6.3in]{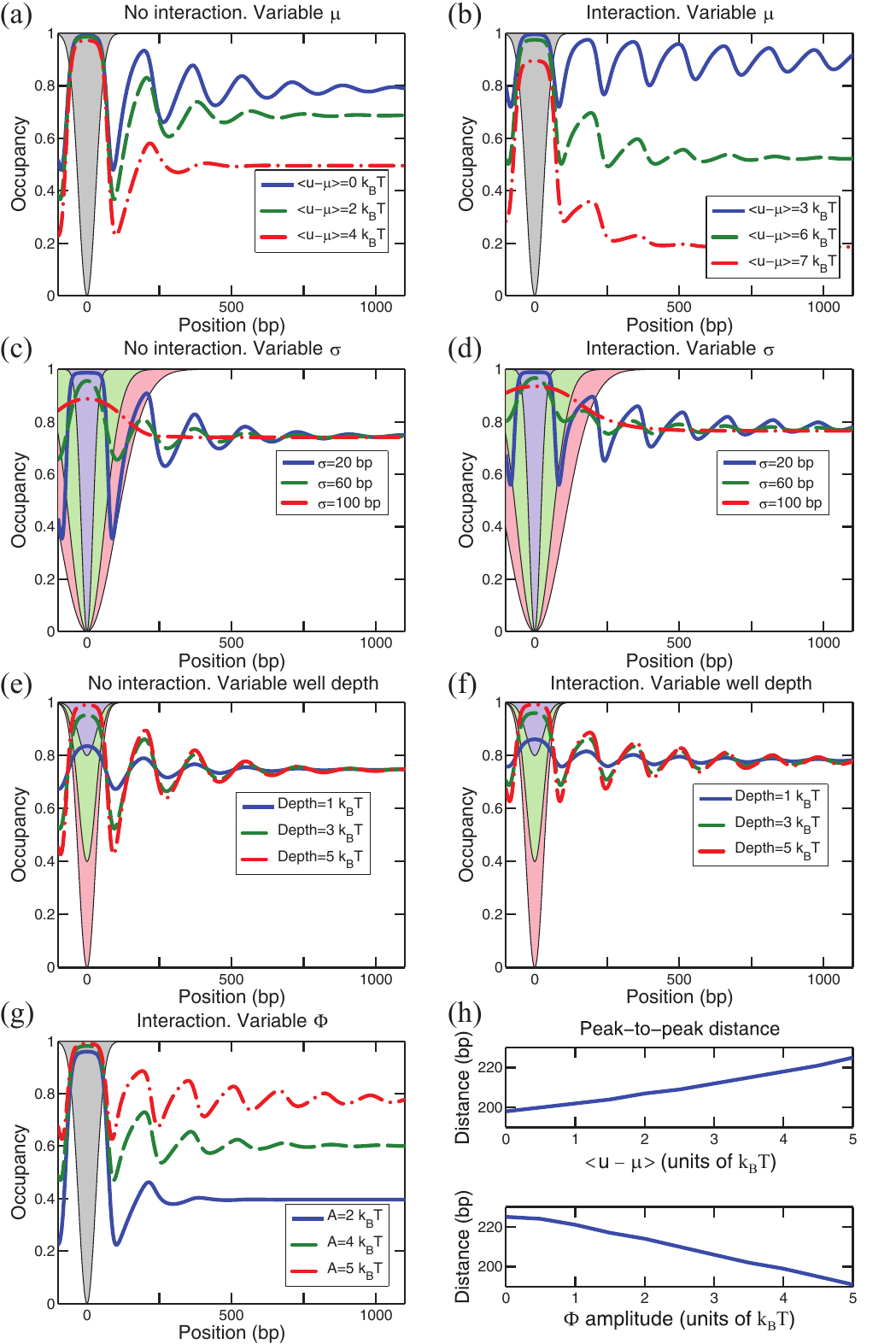}		
  \end{center}
  \caption{(Color online) Symmetric Gaussian well. Occupancy profiles for the scenarios
described in Fig.~\ref{Fig:GaussBarrier}. All the parameters not explicitly given in the legends are from Fig.~\ref{Fig:GaussBarrier}.
In particular, well depths have the same magnitude as the heights of the corresponding barriers.
}
\label{Fig:GaussWell}
\end{figure}

In Fig.~\ref{Fig:GaussBarrier} we show nucleosome occupancy induced by a symmetric Gaussian barrier, with and without two-body interactions.
As the chemical potential is changed to increase the average occupancy, the oscillations become more prominent.
Without two-body interactions, the peak closest to the barrier is always the highest and the occupancy pattern is
a decaying oscillation [Fig.~\ref{Fig:GaussBarrier}(a)]. Strikingly, including $\Phi$ results in a markedly different occupancy profile:
oscillations are more persistent and the peak closest to the barrier is not always the highest [Fig.~\ref{Fig:GaussBarrier}(b)].

The degree of nucleosome localization is also controlled by the width of the Gaussian barrier: wider barriers induce less prominent
oscillations, but produce stronger occupancy depletion over the barrier itself [Figs.~\ref{Fig:GaussBarrier}(c) and \ref{Fig:GaussBarrier}(d)].
The degree of depletion is also controlled by the barrier height [Figs.~\ref{Fig:GaussBarrier}(e) and \ref{Fig:GaussBarrier}(f)].
Interestingly, increasing the strength of two-body interactions results in a higher average occupancy and produces shorter peak-to-peak
distances [Fig.~\ref{Fig:GaussBarrier}(g)]. In fact, the peak-to-peak distances (which can be interpreted as the sum of the 147 bp nucleosomal
site and a linker) can be varied in a wide range by changing either the chemical potential $\mu$ or the strength of $\Phi$  [Fig.~\ref{Fig:GaussBarrier}(h)].

\begin{figure}[t]
  \begin{center}
  \includegraphics[height=6.3in]{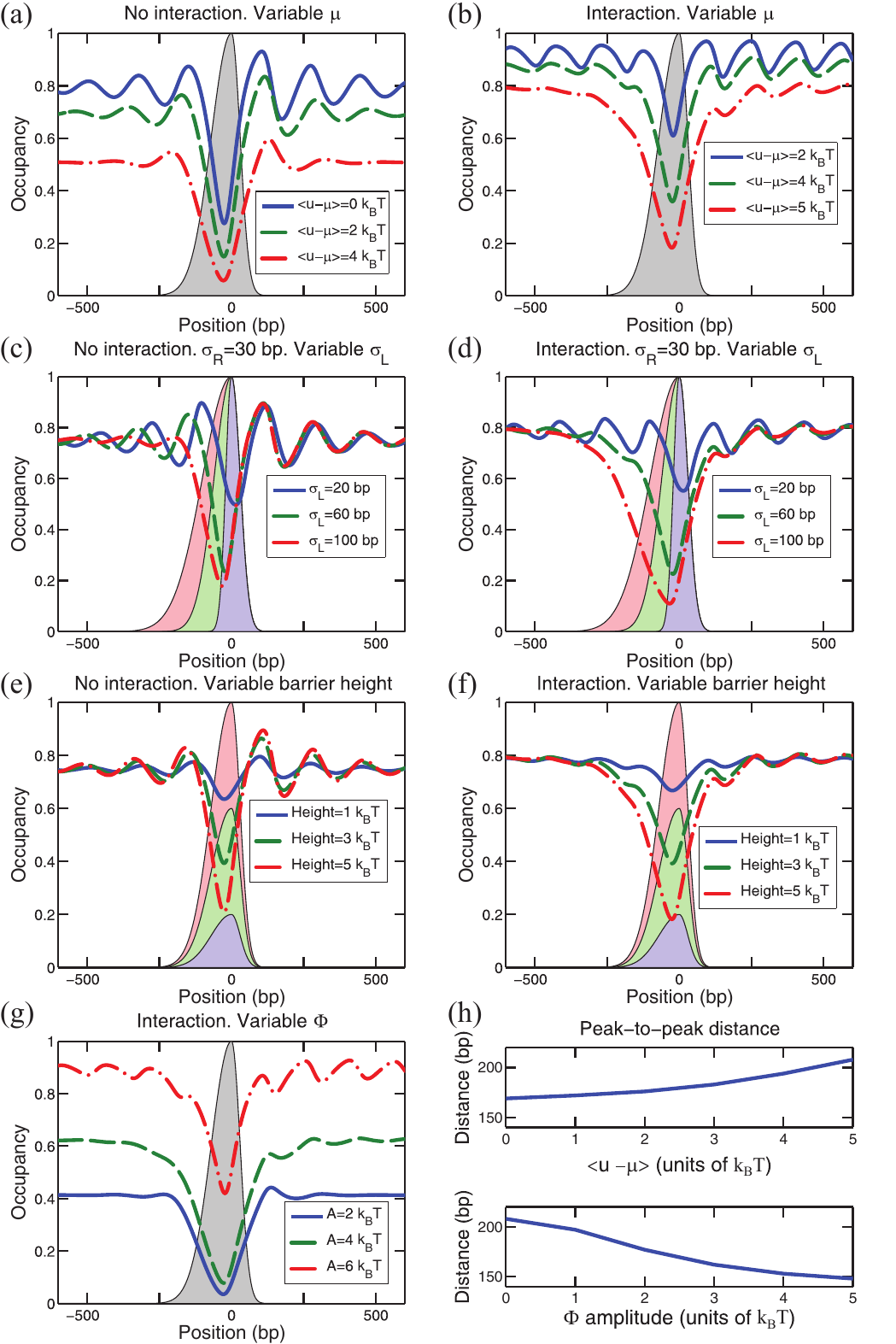}		
  \end{center}
  \caption{(Color online) Asymmetric Gaussian barrier. Occupancy profiles for the scenarios
described in Fig.~\ref{Fig:GaussBarrier}.
Unless otherwise specified in the legend, the barrier heights are $5~k_B T$,
$\sigma_L=70$ bp, $\sigma_R=30$ bp, and $\la u - \mu \ra=5~k_B T$ [in panel (c) $\la u - \mu \ra=1~k_B T$].
}
\label{Fig:AsymGaussBarrier}
\end{figure}
 
Similar conclusions can be reached if the Gaussian barrier is replaced by a symmetric Gaussian potential well (Fig.~\ref{Fig:GaussWell}): oscillations decay less
rapidly with two-body interactions, and the extent of oscillations is controlled by the chemical potential and by the depth and the width of the well. However, in this
case the nucleosome closest to the well is always the most localized.
Nucleosome occupancy in the vicinity of 5' NDRs is prominently asymmetric \cite{Zhang:2009,Mavrich:2008b}.
This asymmetry can be modeled by a combination of a symmetric barrier with an adjacent potential well \cite{Chereji:2010}, or by a single
asymmetric barrier. In Fig.~\ref{Fig:AsymGaussBarrier} we show how nucleosome localization
and the degree of asymmetry in the occupancy profile vary with the chemical potential, the strength of the two-body interaction,
the height of the barrier, and the degree of its asymmetry.

\begin{figure}[!t]
  \begin{center}
  \includegraphics[width=\textwidth]{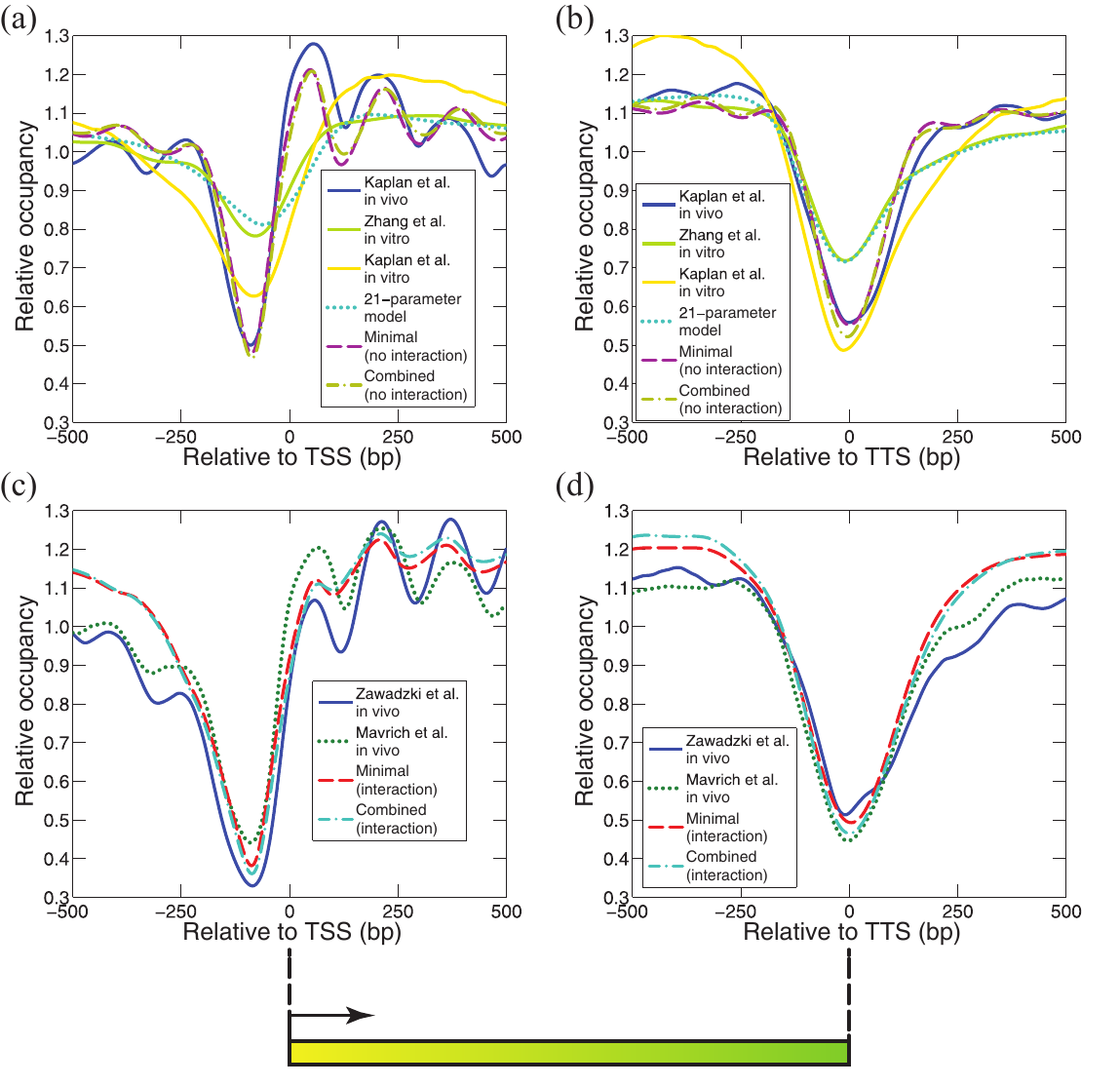}		
  \end{center}
  \caption{(Color online) Average nucleosome occupancy in the vicinity of transcription start and termination sites (TSS and TTS, respectively).
Each occupancy profile is normalized by its average in the $[-500,500]$ bp window.
(a), (b): Nucleosome occupancy observed \textit{in vivo} (YPD medium) and \textit{in vitro} by Kaplan et al. \cite{Kaplan:2009}
and \textit{in vitro} by Zhang et al. \cite{Zhang:2009},
and predicted using a $21$-parameter $N=2$ position-independent model, a minimal model in which nucleosomes are localized purely by means of
sequence-independent potential barriers (Fig.~\ref{Fig:asym_one-body}), and a combined model in which sequence-specific energies
from the $21$-parameter $N=2$ model are added to the barriers from Fig.~\ref{Fig:asym_one-body}.
The two-body potential is turned off. Note that in Ref.~\cite{Zhang:2009} DNA was mixed with histones in a 1:1 mass ratio which is close to the \textit{in vivo} value, while
in Ref.~\cite{Kaplan:2009} the ratio was 0.4:1, resulting in deeper NDRs.
(c), (d): Nucleosome occupancy observed \textit{in vivo} by Zawadzki et al. \cite{Zawadzki:2009} and Mavrich et al. \cite{Mavrich:2008b},
and predicted using the $21$-parameter $N=2$ position-independent model, the minimal model, and the combined model.
The two-body potential is given by $\Phi(\Delta)=A \cos \left( 2\pi \Delta/10\right) \exp(-\Delta/50)$, with $A=5~k_BT$.
}
\label{Fig:TSS_TTS}
\end{figure}

In summary, two-body interactions significantly modify nucleosome occupancy profiles, affecting heights and spacings of
observed nucleosome localization peaks. However, as the interaction itself only couples neighboring nucleosomes but does not determine their absolute positions,
a potential barrier or well is required to achieve localization in the first place. Increasing the width of this feature diminishes its
localization capacity. The interaction favors configurations with linker lengths corresponding to the minima of $\Phi$, leading
to linker length discretization \cite{Widom:1992,Wang:2008,Chereji:2010}.
By changing the interaction strength or the chemical potential one can create occupancy patterns with different average linker lengths.

\subsection{Modeling nucleosome occupancy over transcribed regions} \label{subsection:Results3}

\begin{figure}[t]
  \begin{center}
  \includegraphics[width=4.5in]{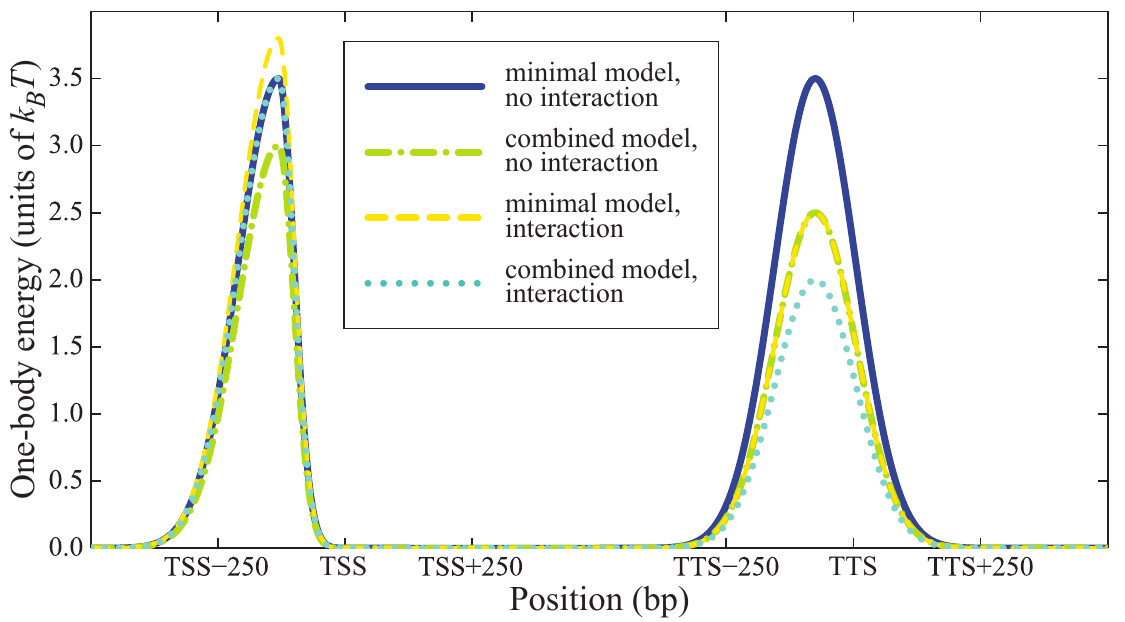}		
  \end{center}
  \caption{(Color online) The one-body energy profiles used in Figs.~\ref{Fig:TSS_TTS} and \ref{Fig:HeatMaps}.
The $5$' asymmetric barrier has $\sigma_\text{left}=80$ bp and $\sigma_\text{right}=30$ bp.
The $3$' symmetric barrier has $\sigma=80$ bp.
Solid blue line: barriers used in the \textit{in vivo} minimal model without two-body interactions [Figs.~\ref{Fig:TSS_TTS}(a) and \ref{Fig:TSS_TTS}(b), Fig.~\ref{Fig:HeatMaps}]. 
Dash-dot green line: barriers used in the \textit{in vivo} combined model without two-body interactions [Figs.~\ref{Fig:TSS_TTS}(a) and \ref{Fig:TSS_TTS}(b), Fig.~\ref{Fig:HeatMaps}].
Dashed yellow line: barriers used in the \textit{in vivo} minimal model with two-body interactions [Figs.~\ref{Fig:TSS_TTS}(c) and \ref{Fig:TSS_TTS}(d)].
Dotted light blue line: barriers used in the \textit{in vivo} combined model with two-body interactions [Figs.~\ref{Fig:TSS_TTS}(c) and \ref{Fig:TSS_TTS}(d)].
The landscapes shown in the Figure are shifted vertically so that $\langle u - \mu \rangle = 0.56~k_BT$ in the minimal model without two-body interactions,
$0.62~k_BT$ in the combined model without two-body interactions, $4.49~k_BT$ in the minimal model with two-body interactions,
and $4.62~k_BT$ in the combined model with two-body interactions.
}
\label{Fig:asym_one-body}
\end{figure}

The characteristic patterns of nucleosome occupancy in the region between $5$' and $3$' NDRs are shown in Fig.~\ref{Fig:TSS_TTS}.
As discussed in the Introduction, there is a pronounced lack of nucleosome localization \textit{in vitro} \cite{Kaplan:2009,Zhang:2009}
[Figs.~\ref{Fig:TSS_TTS}(a) and \ref{Fig:TSS_TTS}(b)]. The $21$-parameter $N=2$ position-independent model
captures this liquid-like behavior correctly but is unable to account for the \textit{in vivo} peaks.
Since DNA sequence alone clearly cannot produce the observed degree of \textit{in vivo} localization,
we sought to construct a minimal model in which potential barriers of non-sequence origin flank each gene and
the one-body energy landscape is flat otherwise \cite{Mobius:2010,Chevereau:2009} (Fig.~\ref{Fig:asym_one-body}).

In the Kaplan et al. dataset \cite{Kaplan:2009}, the first nucleosome is in fact the most localized and the average profile is consistent with the absence of two-body
interactions [Fig.~\ref{Fig:TSS_TTS}(a)]. In contrast, Zawadzki et al. \cite{Zawadzki:2009} and Mavrich et al. \cite{Mavrich:2008b} profiles appear to be shaped by
the higher-order chromatin structure [Fig.~\ref{Fig:TSS_TTS}(c)]. This experimental discrepancy may have resulted from under-digesting
chromatin with micrococcal nuclease (MNase), an enzyme typically used to liberate mononucleosome cores \cite{Weiner:2010}.
In addition, the number of active genes that presumably reside in more open, active chromatin characterized by weaker two-body interactions could vary between experiments. 
However, in all three cases the \textit{in vivo} barriers are necessary to reproduce observed localization patterns. The $5$' NDR is strongly asymmetric
[Figs.~\ref{Fig:TSS_TTS}(a) and \ref{Fig:TSS_TTS}(c)] and thus needs to be modeled either with a combination of a symmetric barrier and
a potential well for the $+1$ nucleosome \cite{Chereji:2010}, or with a single asymmetric barrier (Fig.~\ref{Fig:AsymGaussBarrier}). The height of
each barrier in Fig.~\ref{Fig:asym_one-body} is adjusted to reproduce the extent of observed nucleosome depletion.

The average occupancy profiles are not significantly altered if sequence-specific energies from the 21-parameter $N=2$ position-independent model
are added to the barriers from Fig.~\ref{Fig:asym_one-body} (Fig.~\ref{Fig:TSS_TTS}, cf. combined and minimal models).
The $N=2$ model yields a standard deviation of $0.61~k_B T$ for the energies genome-wide, consistent with the assumption that
sequence-dependent energies should be less than $1~k_B T$, and thus the one-body landscape is still dominated by
the barriers. Note that barrier heights are reduced in the combined model because sequence-dependent nucleosome
depletion over NDRs is now included explicitly.

\begin{figure}[t]
  \begin{center}
  \includegraphics[height=5.6in]{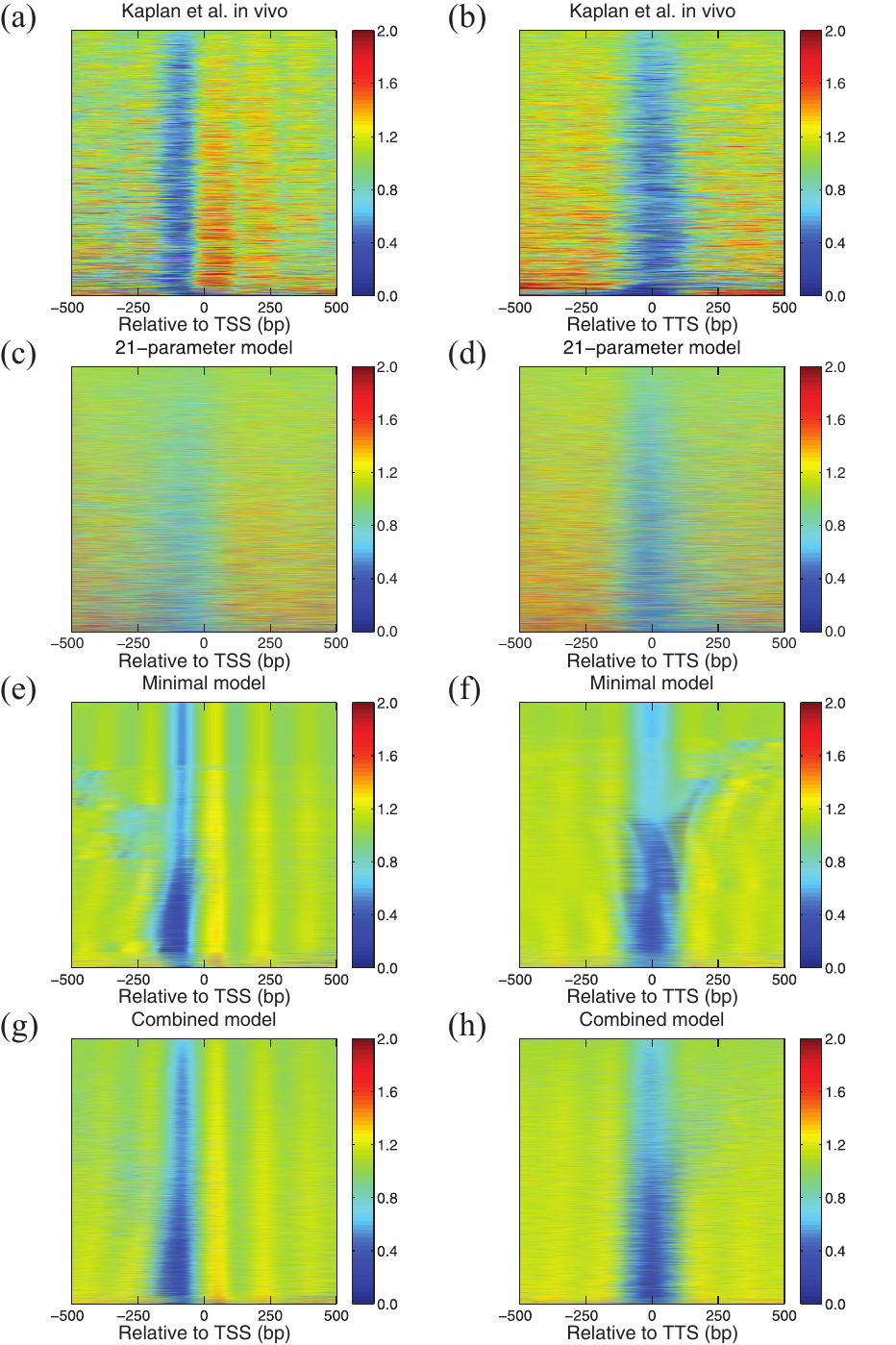}		
  \end{center}
  \caption{(Color online) Heat maps of nucleosome occupancy around TSS and TTS for $5747$ \textit{S. cerevisiae} genes.
\textit{In vivo} nucleosomes (YPD medium) \cite{Kaplan:2009} (a) and (b), $N=2$ position-independent model (c) and (d),
minimal model (e) and (f), combined model (g) and (h).
The minimal model is constructed by placing potential barriers from Fig.~\ref{Fig:asym_one-body} at the end of each gene onto an otherwise
flat one-body energy landscape without two-body interactions.
The combined model is constructed by adding sequence-specific energies from the $21$-parameter $N=2$ position-independent model
(which have standard deviation of $0.61~k_B T$ genome-wide) to the minimal model.
The occupancy for each gene is normalized by the average occupancy in the $[-500,500]$ bp window. The experimental data [(a) and (b)] are smoothed
with a $2D$ Gaussian kernel ($\sigma_X=1$ bp and $\sigma_Y=2$ genes). The genes are sorted in each panel in the order of increasing variance of the occupancy. 
The genome-wide average occupancies are $0.1508$ [(a) and (b)], $0.2024$ [(c) and (d)], $0.7516$ [(e) and (f)], and $0.7232$ [(g) and (h)].
}
\label{Fig:HeatMaps}
\end{figure}

The difference between the minimal and the combined models is more pronounced if individual occupancy profiles are displayed as a heat map (Fig.~\ref{Fig:HeatMaps}).
Minimal model barriers adjacent to each other on the genomic sequence (\textit{e.g.} the $5$' barriers of two divergent genes
sharing a single promoter) sometimes create anomalous NDRs with the extent of nucleosome depletion not observed in the data
[Figs.~\ref{Fig:HeatMaps}(e) and \ref{Fig:HeatMaps}(f)]. Interestingly, these effects are reduced when sequence specificity
is combined with the minimal model [Figs.~\ref{Fig:HeatMaps}(g) and \ref{Fig:HeatMaps}(h)].
Comparing barrier heights in the minimal and combined models (Fig.~\ref{Fig:asym_one-body}), we conclude that intrinsic histone-DNA
interactions are responsible for $< 30 \%$ of the barriers. 

\section{Discussion and Conclusion}

We have developed a theory of nucleosomes positioned by both intrinsic histone-DNA interactions and a two-body, nearest-neighbor potential
induced by higher-order chromatin structure. Our theory is general and can be applied to any system of finite-size particles in an external field
of arbitrary strength, interacting via a short-range two-body potential.
From one- and two-body nucleosome energies we can reconstruct one- and two-particle distribution functions exactly.
Conversely, one- and two-particle distributions available from experiment can be used to predict both intrinsic histone-DNA interactions and the two-body
potential with high accuracy, even in strongly inhomogeneous systems dominated by one-body forces.
Finally, we infer sequence determinants of nucleosome positions through a linear fit
of predicted one-body energies to a model in which each mono- and dinucleotide is assigned the same energy regardless of its position within
the nucleosomal site (the $N=2$ position-independent model). In contrast to our previous approach which employed a hierarchy of energy constraints \cite{Locke:2010},
here we chose to fit all available parameters. The fitted parameters are subsequently projected onto the relevant subspace, creating a non-degenerate description of nucleosome
energetics.

Unfortunately, high-throughput maps of nucleosome
positions typically yield just one-particle distributions -- two-body information is lost because the positional data is averaged over many cells.
Nonetheless, the two-body potential can be inferred from one-body density profile $n(i)$ if $\Phi$ is sufficiently
strong to create $10-11$ bp periodic occupancy oscillations in the vicinity of potential wells and barriers which tend to localize nucleosomes (Fig.~\ref{Fig:Model}).
These oscillations can then be used to predict $P_\text{linker}$, the distribution of linker lengths, which gives $\Phi$ via Eq.~\eqref{Eq:Phi:est}.
This empirical approach fails in the absence of one-body energies because arrays of nucleosomes (even subject to two-body forces)
can freely move along the DNA without any energy cost, resulting in a constant $n(i)$ from which $P_\text{linker}$ cannot be deduced.

We have previously shown that our approach is capable
of separating the $10-11$ bp periodic two-body potential from the one-body contribution which
is responsible for rotational nucleosome positioning and thus has the same periodicity of DNA helical twist \cite{Chereji:2010}.
Furthermore, we have inferred two-body interactions from genome-wide maps of nucleosomes assembled \textit{in vitro} on genomic DNA,
and have demonstrated their essential role in explaining the observed autocorrelation of genome-wide nucleosome positions, and in shaping the occupancy
profile over transcribed regions.

However, the full analysis, which involves finding $u(i)$ from a system of nonlinear
equations obtained by inserting $\Phi$ from Eq.~\eqref{Eq:Phi:est} into Eq.~\eqref{Eq:n}, is computationally expensive.
A simpler approach which neglects two-body interactions [Eq.~\eqref{Eq:u0}] is much faster but leads to ``leakage'' of two-body effects into
predicted one-body potentials if $\Phi$ is in fact present in the system (Fig.~\ref{Fig:Model}).
Nonetheless, if a sequence-specific model is fit to this one-body profile, the correct sequence-dependent energy is obtained
because oscillations caused by $\Phi$ are not related to any sequence features. Thus if one only needs to determine sequence determinants of nucleosome
positioning, the two-body potential can be neglected altogether and a simpler, computationally efficient approach can be used.
Nevertheless, only the full theory presented here is capable of disentangling one- and two-body effects.

\textit{In vivo}, nucleosomes adjacent to $5$' and $3$' NDRs are positioned by steric exclusion which creates prominent
localized peaks of nucleosome occupancy in the vicinity of each depleted region \cite{Kaplan:2009,Yuan:2005,Mavrich:2008a,Mavrich:2008b}. 
This localization is absent \textit{in vitro} \cite{Kaplan:2009,Zhang:2009}, indicating that intrinsic histone-DNA interactions can only partially
account for \textit{in vivo} occupancy patterns, creating barriers with heights less than $1~k_B T$. Therefore, sequence-specific models need to be
combined \textit{in vivo} with additional potentials established by RNA polymerase, ATP-dependent chromatin remodeling enzymes,
and DNA-binding proteins.

To this end, we have studied nucleosome positioning by potential wells and barriers, showing in particular that
the distance between neighboring peaks in the nucleosome occupancy profile depends on both histone octamer concentration
and the strength of two-body interactions (Figs.~\ref{Fig:GaussBarrier}, \ref{Fig:GaussWell} and \ref{Fig:AsymGaussBarrier}).
We then created a genome-wide model of nucleosome occupancy by flanking each transcribed region
with potential barriers whose height, width and the degree of asymmetry were adjusted to reproduce the observed occupancy patterns (Fig.~\ref{Fig:asym_one-body}).
This minimal, sequence-independent model successfully captures the main features of \textit{in vivo} occupancy, although, surprisingly, the role of $\Phi$ seems
to vary between experiments (Fig.~\ref{Fig:TSS_TTS}).
When the minimal model is combined with the sequence-specific energies (with magnitudes $\le 1~k_B T$ as discussed above),
occupancy patterns of a subset of genes get closer to experiment (Fig.~\ref{Fig:HeatMaps}),
but the effect of DNA sequence on the average occupancy profile is rather small (Fig.~\ref{Fig:TSS_TTS}).

In summary, we have presented a rigorous theory of nucleosomes subject to nearest-neighbor two-body interactions, demonstrating
the essential role of chromatin structure in genome-wide models of nucleosome occupancy. Future experiments
focused on multi-nucleosome distributions will provide data for our exact theory [Eqs.~\eqref{Eq:u} and \eqref{Eq:Phi}], obviating
the need for the approximate Eq.~\eqref{Eq:Phi:est}. Further experimental and computational work should also be aimed at pinpointing the origin of \textit{in vivo}
potential barriers in various organisms and cell types.

\begin{acknowledgements}
This research was supported by National Institutes of Health (HG004708)
and by an Alfred P. Sloan Research Fellowship to A.V.M.
\end{acknowledgements}

\bibliographystyle{spphys}         

\end{document}